\documentclass[twocolumn,floatfix,superscriptaddress,nofootinbib]{revtex4}
\usepackage{graphicx}
\usepackage{amsmath}
\usepackage{amssymb}
\usepackage{booktabs}
\usepackage{color}
\usepackage{float}
\usepackage[utf8]{inputenc}

\usepackage[]{natbib}

\newcommand{\be}{\begin{eqnarray}}
\newcommand{\ee}{\end{eqnarray}}
\newcommand{\bse}{\begin{subequations}}
\newcommand{\ese}{\end{subequations}}

\newcommand{\bnum}{\begin{enumerate}}
\newcommand{\enum}{\end{enumerate}}

\newcommand{\bit}{\begin{itemize}}
\newcommand{\eit}{\end{itemize}}

\newcommand{\bc}{\begin{cases}}
\newcommand{\ec}{\end{cases}}

\newcommand{\bpm}{\begin{pmatrix}}
\newcommand{\epm}{\end{pmatrix}}

\newcommand{\bvm}{\begin{vmatrix}}
\newcommand{\evm}{\end{vmatrix}}

\newcommand{\bs}{\boldsymbol}

\newcommand{\mcal}{\mathcal}

\newcommand{\gc}{\gamma}
\newcommand{\gd}{\delta}
\newcommand{\eps}{\epsilon}

\newcommand{\gl}{\lambda}

\newcommand{\go}{\omega}

\newcommand{\Gl}{\Lambda}

\newcommand{\p}{\partial}
\newcommand{\f}{\frac}

\newcommand{\csp}{\;,\qquad}

\newcommand{\markblue}[1]{\textcolor{black}{#1}}
\newcommand{\markblack}[1]{\textcolor{black}{#1}}

\begin{document}


\title{Antipolar ordering of topological defects in active liquid crystals}

\author{Anand U. Oza} 
\affiliation{Courant Institute of Mathematical Sciences,
251 Mercer Street,  New York, NY 10012, USA}

\author{J\"orn Dunkel}
\affiliation{Department of Mathematics, Massachussetts Institute of Technology, 77 Massachusetts Avenue, Cambridge, MA 02139-4307, USA}
\date{\today}
   
\begin{abstract}
ATP-driven microtubule-kinesin bundles can self-assemble into two-dimensional active liquid crystals (ALCs) that exhibit a rich creation and annihilation dynamics of topological defects, reminiscent of particle-pair production processes in quantum systems. This recent discovery has sparked considerable interest but a quantitative theoretical description is still lacking. We present and validate a minimal continuum theory for this new class of active matter systems by generalizing the classical Landau-de Gennes free-energy to account for the experimentally observed spontaneous buckling of motor-driven extensile microtubule bundles. The resulting model agrees with recently published data and predicts a regime of antipolar order. Our analysis implies that ALCs are governed by the same generic ordering principles that determine the non-equilibrium dynamics of dense bacterial suspensions and elastic bilayer materials. Moreover, the theory manifests an energetic analogy with strongly interacting quantum gases. Generally, our results suggest that complex non-equilibrium pattern-formation phenomena might be predictable from a few fundamental symmetry-breaking and scale-selection principles.
 \end{abstract}
 

\maketitle

Active materials~\cite{2015McEvoy} assembled from intracellular components, such as DNA~\cite{2009Douglas_NuclAcidsRes}, microtubules and motor proteins~\cite{2010Bausch,2012Sumino,2012Sanchez_Nature}  promise innovative biotechnological applications, from microscale transport and medical devices~\cite{2009Douglas_NuclAcidsRes}  to artificial tissues~\cite{2015McEvoy} and programmable soft materials~\cite{2006Rothemund_Nature,2014Nickels_Small,2014Studart}. Beyond their practical value, these systems challenge theorists to generalize equilibrium statistical mechanics to far-from-equilibrium regimes~\cite{1998TonerTu_PRE,2002Ra,2008SaintillanShelley,2011Maha_PRL,2011Peruani_JPhysConf,2012Peshkov,2012Wensink,2012Wensink_JPhysCM,2013Ravnik_PRL,2013Adhyapak_Soap,2013Marchetti_Review,PhysRevLett.114.048101,2015Giomi,2015Krieger,2015Trivedi,2016Menzel,2016Putzig_SoftMatter,2016Yeomans_NatComm}. Recent experimental advances in the self-assembly and manipulation of colloids with DNA-mediated interactions~\cite{Mirkin:1996aa,2005Valignat_PNAS,2013Palacci} have stimulated theoretical analysis that may eventually help clarify the physical principles underlying self-replication~\cite{Ebeling,2014Zeravcic_PNAS,2013England} and evolution in viruses~\cite{Berger:1994aa,2015PerlmutterHagan,2014Meng_Science} and other basic biological systems. Yet, despite some  partial progress~\cite{2002Ra,2008BaMa,2009BaMa_PNAS,2013Marchetti_Review,2012Vicsek,2010Ramaswamy,2013Adhyapak_Soap}, our conceptual understanding of active materials, and living matter in general, remains far from complete. We do not know whether, or under which conditions,  \lq universality\rq{} ideas~\cite{Goldenfeld:2011aa} that have proved powerful in the description of equilibrium systems can be generalized to describe collective dynamics of active matter not just qualitatively but also quantitatively.  This deficit may be ascribed to the fact that mathematical models have been successfully tested against experiments in only a few instances~\cite{2010Bausch,2012Sumino,2012Wensink,2013Dunkel_PRL,PhysRevLett.110.208001,2015Bausch_NatPhys}.
\par

Recently discovered 2D active liquid crystal (ALC) analogs~\cite{2007NaRaMe,2008Aranson,2012Sanchez_Nature,Shi:2013aa,2014Keber_Science} comprise an important class of non-equilibrium systems that allows further tests of general theoretical concepts~\cite{Goldenfeld:2011aa} and specific models. ALCs are  assemblies of rod-like particles that exhibit non-thermal collective excitations due to steady external~\cite{2007NaRaMe,2008Aranson} or internal~\cite{2012Sanchez_Nature,2014Keber_Science} energy input. At high concentrations, ALCs form an active nematic phase characterized by dynamic creation and annihilation of topological defects~\cite{2007NaRaMe,2012Sanchez_Nature,2014Keber_Science}, reminiscent of spontaneous particle-pair production in quantum systems. This phenomenon was demonstrated recently~\cite{2012Sanchez_Nature,2014Keber_Science,2015DeCamp} for ATP-driven microtubule-kinesin bundles trapped in flat and curved  interfaces.   Moreover, these experiments~\cite{2015DeCamp} revealed an unexpected nematic ordering of topological defects which is unaccounted for in current theoretical models.  Understanding the emergence of such topological super-structures is crucial for the development and control of new materials, as recently demonstrated for colloidal liquid crystals~\cite{2006Musevic_Science,Machon27082013,Tkalec01072011}.

\par
We here develop and test a closed continuum theory for dense ALCs by generalizing the higher-order scalar and vector theories of soft elastic materials~\cite{2015Stoop_NatMat} and bacterial fluids \cite{2012Wensink,2013Dunkel_PRL} to matrix-valued fields. Specifically, we propose a modification of the commonly-adopted Landau-de Gennes (LdG) free energy to account for the inherently different microscopic buckling behaviors of passive and active LCs~\cite{2012Sanchez_Nature}. While bending is energetically unfavorable in passive LCs and hence penalized by the LdG energy functional, kinesin-driven  ALCs buckle spontaneously even at low concentrations due to the extensile motor action (Fig. 1\textbf{a}). This experimental observation~\cite{2012Sanchez_Nature} implies that the classical LdG framework is, by construction, ill-suited to describe experiments  in which microtubule bundles are sheared relative to each other by motor proteins~\cite{2012Sanchez_Nature,2014Keber_Science,2015DeCamp}. The inclusion of active bending effects in the LdG functional yields a tensor version of the Swift-Hohenberg theory~\cite{1977SwiftHohenberg} of pattern formation. The resulting minimal model has only two dimensionless parameters, thus allowing a detailed comparison with recent experimental data~\cite{2012Sanchez_Nature,2015DeCamp}.

\par
In addition to the traditional $Q$-tensor formulation, we present an equivalent complex scalar field representation~\cite{PhysRevA.38.2132,2012Peshkov}  that manifests an analogy with a generalized Gross-Pitaevskii theory~\cite{1961Gross,1961Pitaevskii} of strongly coupled many-body quantum systems~\cite{PhysRevLett.88.090401,Lin:2011aa,PhysRevLett.109.095302,Parker:2013aa}. In the case of normal dispersion, the celebrated LC-superconductor correspondence~\cite{1972deGennes,PhysRevA.38.2132} has helped elucidate profound parallels between the smectic phase in passive LCs and the Abrikosov vortex lattices in type-II superconductors~\cite{1989Goodby_Nature,1990Srajer_PRL}.  The results below indicate that a similar  analogy may exist between ALCs and Bose-Einstein/Fermi condensates with double-well dispersion~\cite{Lin:2011aa,Parker:2013aa,PhysRevLett.109.095302}, suggesting that ALCs could offer insights into the dynamics of these quantum systems and \textit{vice versa.}

\section*{RESULTS}

\begin{figure*}[t]
\includegraphics[width=0.85\textwidth]{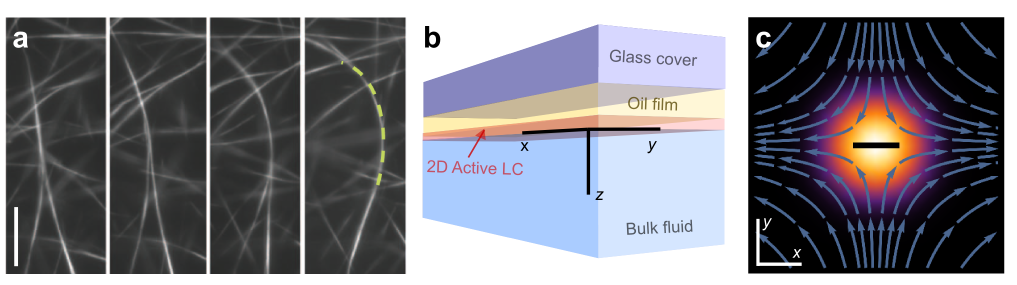}
\caption{(\textbf{a})~\markblack{Image sequence showing spontaneous buckling of a microtubule bundle (dashed) caused by extensile ATP-driven motor activity, adapted with permission from Fig.~1\textbf{c} in Ref.~\cite{2012Sanchez_Nature}. Time interval~$11.5\,$s, scale bar~$15\,\mu$m. For a passive or contractile bundle, one would expect straightening instead of bending, approximately corresponding to a time-reversal of the depicted sequence.}
(\textbf{b})~Schematic of the experimental setup reported in Ref.~\cite{2012Sanchez_Nature,2015DeCamp}, not drawn to scale. A thin oil film (thickness~$\sim 3\,\mu$m) separates a 2D ALC layer ($\sim 0.2-1.0\,\mu$m) at the oil-water interface from a solid glass cover. Liquid can be exchanged between the ALC layer and bulk fluid, resulting in compressible 2D interfacial flow that is strongly damped by the nearby no-slip glass boundary and the viscous oil layer. 
(\textbf{c})~Extensile 2D dipole flow in the interface as predicted by the overdamped closure condition~\eqref{e:closure} for $D>0$ and $Q=(\gl,0;0,-\gl)$ with $\gl=\exp(-\bs r^2)$. The central horizontal bar indicates the unit director axis,  and background colors the nematic order parameter~\mbox{$S\sim\gl$}.
\label{fig:closure}
}
\end{figure*}

\textbf{Experimental conditions.}
Recent experiments~\cite{2012Sanchez_Nature,2015DeCamp} show that  ATP-driven microtubule-kinesin bundles can self-assemble into a dense quasi-2D ALC layer at a surfactant-supported oil-water interface parallel to a planar solid boundary~(Fig.~\ref{fig:closure}\textbf{b}). This \lq wet\rq{} ALC was found to exhibit local nematic alignment of bundles, persistent annihilation and creation dynamics of topological defects~\cite{2012Sanchez_Nature}, and remarkable nematic order of the defect orientations in thin layers~\cite{2015DeCamp}.  Although a large number of unknown parameters has prevented detailed quantitative comparisons between theory and experiment, several recently proposed multi-order-parameter models of 2D ALC systems~\cite{2013Thampi_PRL,2013Giomi_PRL,2012Peshkov,PhysRevLett.114.048101} were able to reproduce qualitatively selected aspects of these observations, such as defect-pair creation and separation~\cite{2013Giomi_PRL}. Despite providing some important insights, traditional models often do not account for three relevant details of the experiments~\cite{2012Sanchez_Nature,2015DeCamp}. First, those models typically assume divergence-free 2D fluid flow within the ALC layer, which is a valid approximation for isolated free-standing film experiments~\cite{2007SoEtAl} but neglects fluid exchange between the 2D interface and bulk in the ALC experiments (Fig.~\ref{fig:closure}\textbf{b}). \markblack{Indeed, the surfactant-stabilized interface causes the microtubule-kinesin bundles to assemble into a quasi-2D layer, but places no such constraint on the fluid.} As is known for classical turbulence~\cite{1980Kraichnan,2002Kellay},  small-scale energy input can trigger turbulent upward cascades in incompressible 2D flow.  Thus, topological defect dynamics in the current standard models may be dominated by artificially enhanced hydrodynamic mixing due to a simplifying 2D incompressibility assumption that is unlikely to hold under realistic  experimental conditions~\cite{2012Sanchez_Nature,2015DeCamp}. Second, a relevant yet previously ignored effect is damping from the nearby boundaries, which may promote topological defect ordering. \markblack{Third, as already mentioned above, the commonly adopted standard LdG free-energy functional does \emph{not} account for motor-driven spontaneous buckling~\cite{2012Sanchez_Nature} of microtubule bundles (Fig.~\ref{fig:closure}\textbf{a}), which is one of the key differences between passive and active LCs (Z. Dogic, private communication).} To overcome such limitations and achieve a quantitative description of the experiments~\cite{2012Sanchez_Nature,2015DeCamp}, we next construct a closed continuum theory for ALCs described by a nematic tensor field  $Q(t,\bs r)$. The theory accounts for the different buckling behaviors of passive and active LCs and builds on a self-consistent hydrodynamic closure condition. 

\textbf{Theory.}
Traditional multi-field models~\cite{2013Thampi_PRL,2013Giomi_PRL} aim to describe the  2D  nematic phase of a dense ALC suspension by coupling  the  filament concentration $c(t,\bs r)$ and  the nematic order tensor $Q(t,\bs r)$ to an incompressible 2D flow field $\bs v(t,\bs r)$ that satisfies \mbox{$\nabla\cdot\bs v=0$} in the interface plane $\bs r=(x,y)$. The nematic order parameter $S(t,\bs r)$ is proportional to the larger eigenvalue of $Q$, and the filaments are oriented along the corresponding eigenvector, or director $\bs n(t,\bs r)$. To construct an alternative closed-form theory for the symmetric traceless $2\times 2$-tensor field~$Q$,  we start from the generic transport law
\be\label{e:Q-dynamics}
\p_tQ+\nabla\cdot (\bs v Q)-\kappa[Q,\go]=-\f{\gd\mcal{F}}{\gd Q}
\ee
where $\go=[\nabla \bs v-(\nabla \bs v)^\top]/2$ is the vorticity tensor,  $[A,B] = AB-BA$ the commutator of two matrices and $\mcal{F}[Q]=\int d^2r\, F$ an effective free energy. Focussing on dense suspensions as realized in the experiments~\cite{2012Sanchez_Nature,2015DeCamp}, we neglect fluctuations in the microtubule concentration, $\nabla c\equiv 0$. A derivation of the advection term $\nabla\cdot(\bs v Q)$ from the probability conservation laws underlying generic advection-diffusion models is outlined in the Supplementary Information. It is important, however, that \mbox{$\nabla\cdot (\bs v Q)\ne \bs v\cdot \nabla  Q$} when $\nabla\cdot\bs v\ne 0$,  which is typically the case when fluid can enter and leave the interface.  Combining LdG theory~\cite{1995deGennes} with Swift-Hohenberg theory~\cite{1977SwiftHohenberg}, we consider the \markblue{effective non-equilibrium free-energy density} (Supplementary Information)
\be\label{e:free-energy}
F = 
\text{Tr}\left\{-\frac{a}{2}Q^2+\frac{b}{4}Q^4
-\frac{\gamma_2}{2}(\nabla Q)^2+\frac{\gamma_4}{4}(\nabla\nabla Q)^2\right\}
\ee
with  $a,b>0$ for the nematic phase. Assuming $\gc_2$ can have either sign, ultraviolet stability requires $\gc_4>0$.  For $\gc_2<0$, $F$ penalizes bending and buckling, as is appropriate for passive LCs and possibly \lq dry\rq{} shaken nematics~\cite{2007NaRaMe,2008Aranson},  forcing the system dynamics towards a homogeneous nematic ground-state manifold. By contrast, motor-induced spontaneous buckling~\cite{2012Sanchez_Nature} (Fig.~\ref{fig:closure}\textbf{a}) of kinesin-driven ALCs demands $\gc_2>0$, and consequently patterns of characteristic wavelength $\Gl\sim \sqrt{\gc_4/\gc_2}$ become energetically favorable, as shown below.   
\par

The requirement $\gamma_2 > 0$ for ALCs has an intrinsically microscopic origin, as the ALC assembly consists of microtubules that grow against each other and spontaneously buckle due to the motor-induced extensile shear dynamics of adjacent bundles (Fig.~\ref{fig:closure}{\bf a}). To explain this important point in more detail, let us recall that passive LCs are modeled using $\gamma_2 < 0$, as the corresponding term in the free energy penalizes variations in $Q$ and thus inhibits spatial inhomogeneities at damping rate $\gc_2k^2<0$ in Fourier space. Microscopically, the alignment dynamics of two rod-like passive LC molecules roughly corresponds to a time-reversal of the ALC pair-interaction image sequence in Fig.~\ref{fig:closure}{\bf a}, implying that a corresponding ALC systems would develop buckling instabilities at growth rate $\gc_2k^2>0$. Additional empirical support for this theoretical picture comes from a comparison of the experimentally observed length scales in ALC systems: The microtubule-kinesin bundles realized in the ALC experiments~\cite{2012Sanchez_Nature,2015DeCamp} are approximately 10-30$\,\mu$m in length (Fig.~1{\bf b} in~\cite{2012Sanchez_Nature}), which is on the same scale as both the spontaneous buckling wavelength (Fig.~1{\bf c}, {\bf d} in~\cite{2012Sanchez_Nature}, and Fig.~\ref{fig:closure}{\bf a}) and the typical separation distance between defects (bottom panels of Fig.~3{\bf d} in~\cite{2012Sanchez_Nature}, reproduced in Fig.~\ref{fig:pair}{\bf a} below). That is, there exists no substantial scale separation between the buckling microscopic constituents and emergent ALC dynamics. 
\par
\markblue{Similar buckling phenomena are generically observed in many systems that are subjected to external or internal stresses, for example in elastic films and sheets~\cite{2015Stoop_NatMat,2014Vetter} and in geometrically confined cellular networks~\cite{2014Kang,2014Vaziri}. The ALCs experience an effective compressive stress due to the extensile 'growth' of the filament pairs in a confined geometry, which arises from their motor-induced shear dynamics. Application of such a compressive stress leads to buckling of the network's constituents~\cite{2014Kang,2014Vaziri}. It has been shown that out-of-plane buckling of an elastic sheet due to an effective compressive stress may be quantitatively modeled by a Swift-Hohenberg-type equation with $\gamma_2 > 0$ in the corresponding free energy~\cite{2015Stoop_NatMat}. We expect the same to be true for the in-plane buckling of microtubules confined to a planar interface, and  
thus analyze here an effective theory that incorporates this spontaneous motor-induced buckling phenomenologically through $\gamma_2 > 0$.}
\par 
\textbf{Hydrodynamic closure.}
To obtain a closed $Q$-model, we relate the 2D flow field $\bs v$ to $Q$ through the linearly damped Stokes equation~\markblack{\cite{1988Evans_JFM,2013Woodhouse_PNAS}}
\be\label{e:Stokes} 
-\eta\nabla^2 \bs v+\nu \bs v = -\zeta\nabla\cdot Q
\ee
where $\eta$ is the viscosity and the rhs. represents active stresses~\cite{2002Ra,2013Thampi_PRL} with $\zeta>0$ for extensile ALCs~(Fig.~\ref{fig:closure}\textbf{c}). A pressure term does not appear in Eq.~(\ref{e:Stokes}) because the interfacial flow is not assumed to be incompressible and concentration fluctuations are neglected. \markblack{The $\nu$-term in the force balance~\eqref{e:Stokes} has been used to model interfacial damping in other contexts, such as surfactant membranes on a solid substrate~\cite{1988Evans_JFM}}, and  accounts for friction from the nearby no-slip boundary in the Hele-Shaw~\cite{2013Woodhouse_PNAS} approximation  (Fig.~\ref{fig:closure}\textbf{b}). 
In the overdamped regime $\nu \Lambda^2/\eta\gg 1$, we deduce from Eq.~\eqref{e:Stokes} the closure condition 
\be\label{e:closure}
\bs v = -D\nabla\cdot Q  
\csp D=\zeta/\nu.
\ee
 Equation~\eqref{e:closure} is conceptually similar to closure conditions proposed previously for active polar films \cite{VJProst2006}. Importantly,  Eq.~\eqref{e:closure} predicts divergent interfacial flow,~\mbox{$\nabla\cdot\bs v\ne0$}, and hence fluid transport perpendicular to the interface wherever $\nabla\nabla:Q\ne 0$. Inserting~\eqref{e:closure} into~\eqref{e:Q-dynamics} yields a closed $Q$-theory in which periodic director patterns corresponding to local minima of the free energy $\mcal{F}$ can become mixed by self-generated interfacial flow.

\textbf{Complex representation \& ALC-quantum analogy.}
The traditional characterization of 2D nematic order in terms of the symmetric traceless 2$\times 2$ matrix field $Q=(\gl,\mu;\mu,-\gl)$ is redundant, for only two real scalar fields $\gl(t,\bs r)$ and $\mu(t,\bs r)$ are needed to specify  the nematic state at each position $\bs r=(x,y)$. To obtain an irreducible representation~\cite{PhysRevA.38.2132,2012Peshkov} we define the complex position coordinate \mbox{$z=x+iy$},  velocity field \mbox{$v(t,z)=u+iw$} and complex order parameter \mbox{$\psi(t,z)=\gl+i\mu$}, such that $S=2|\psi|$. In terms of the Wirtinger gradient operator \mbox{$\partial=\f{1}{2}(\p_x-i\p_y)$}, the 2D Laplacian takes the form \mbox{$\nabla^2=4\bar{\p}\p$} and the closure condition~\eqref{e:closure} reduces to~\mbox{$v=-2D\partial \psi$}. Denoting the real and imaginary parts of an operator $\mcal{O}$ by $\Re\{\mcal{O}\}$ and $\Im\{\mcal{O}\}$, Eqs.~\eqref{e:Q-dynamics} and~\eqref{e:free-energy} may be equivalently expressed as
\be\label{e:Q-dynamics-complex}
\p_t \psi + \mcal{A}_\psi \psi =-\f{\gd \mcal{G}}{\gd \bar{\psi}}
\ee
where the self-advection operator is given by
\be\label{e:advection-complex}
\mcal{A}_\psi =-4D\,\Re\{(\p^2\psi)+(\p\psi)\p \} +4\kappa D\,i\,\Im\{\partial^2\psi\} 
\ee
and the free energy  $\mcal{G}[\psi,\bar{\psi}]=\int dz\; G$ has the density
\be\label{e:free_energy-complex}
G = 
- a|\psi |^2+\frac{b}{2}|\psi|^4 
+ \gamma_2\bar{\psi}(4\bar{\p}\p) \psi 
+  \gamma_4\bar{\psi}(4\bar{\p}\p)^2\psi.
\ee
For $\gamma_2<0$ and $\gamma_4\to 0$, Eq.~\eqref{e:free_energy-complex} reduces to the energy density of the Gross-Pitaevskii mean-field  model~\cite{1961Gross,1961Pitaevskii} for weakly interacting boson gases. Historically, this limit case has been crucial~\cite{1972deGennes,PhysRevA.38.2132} for elucidating the analogy between the smectic phase of passive LCs and the Abrikosov flux lattice in type-II superconductors \cite{1989Goodby_Nature,1990Srajer_PRL}. For $\gamma_2,\gamma_4>0$, Eq.~\eqref{e:free_energy-complex} effectively describes  double-well dispersion~\cite{PhysRevLett.88.090401}, as recently realized for quasi-momenta in spin-orbit-coupled Bose-Einstein condensates~\cite{Lin:2011aa,Parker:2013aa} and Fermi gases~\cite{PhysRevLett.109.095302}. This fact establishes an interesting connection between dense ALCs and strongly coupled quantum systems: when self-advection is negligible $(D\to 0)$, the fixed point configurations of Eq.~\eqref{e:Q-dynamics-complex} coincide with the \lq eigenstates\rq{} of generalized Gross-Pitaevskii models that incorporate wavelength selection. 

\textbf{Stability analysis.}
The qualitative model dynamics is not significantly altered for moderate values of $\kappa$ (Movies S1 and S2), so we neglect the commutator term by setting $\kappa=0$ from now on (see Supplementary Information for $\kappa>0$).
To understand the properties of Eqs.~\eqref{e:Q-dynamics-complex}--\eqref{e:advection-complex} when self-advection is relevant, we perform a fixed point analysis of the rescaled dimensionless equation (Supplementary Information) 
\be
\label{e:dimless}
&&\p_t \psi -4D\,\Re\{(\p^2\psi)+(\p\psi)\p \}\, \psi=
\notag\\
&&\qquad\qquad
\left(\f{1}{4}-|\psi|^2\right)\psi 
- \gamma_2(4\bar{\p}\p) \psi 
-  (4\bar{\p}\p)^2\psi, 
\quad
\ee 
by focussing on the uniform state $\psi_* = \frac{1}{2}e^{i2\theta}$, which corresponds to a nematic order parameter value \mbox{$S=1$} and homogeneous director angle $\theta$ relative to the $x$-axis. Considering wave-like perturbations \mbox{$\psi=\psi_*+\hat{\eps}(t) e^{i{\bs k}\cdot{\bs r}}
$} with $|\hat{\eps}|\ll 1$ and extensile ALCs with $D>0$, one finds that $\psi_*$ is unstable when $\gc_2>0$ (Supplementary Information).  For subcritical self-advection, \mbox{$D<D_c=2(-\gamma_2+\sqrt{\gamma_2^2+2})$}, the dominant instability is driven by modes with wavenumber $|\bs{k}| = \sqrt{\gamma_2/(2\gamma_4)}$, suggesting the formation of stripe patterns with typical wavelength $\Gl\approx \sqrt{8\pi^2\gc_4/\gc_2}$. By contrast, for supercritical advection,~\mbox{$D>D_c$}, the most unstable mode propagates perpendicular to the director, \mbox{$(k_*,\phi_*)=\left(\sqrt{(2\gamma_2+D)/(4\gc_4)} ,\theta+(\pi/{2})\right)$}, suggesting the possibility of transverse mixing.

\begin{figure}[t!]
\includegraphics[width=0.95\columnwidth]{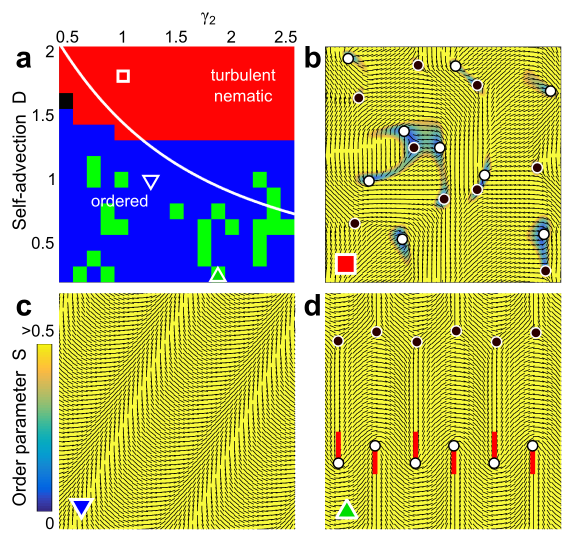}
\caption{
Phase diagram obtained from simulations of Eq.~\eqref{e:dimless} for one particular set of random initial conditions showing the emergence of turbulent nematic states for supercritical active self-advection. (\textbf{a})~We observe convergence to defect-free stripes (blue, panel~\textbf{c}, Movie~S3), long-lived static and oscillatory defect lattice solutions (green, panel~\textbf{d}, Movie~S4), oscillatory defect creation and annihilation events (black, Movies S6 and S7), and chaotic dynamics (red, panel~\textbf{b}, Movie~S1). The white line indicates the analytical estimate ~\mbox{$D_c=2(-\gamma_2+\sqrt{\gamma_2^2+2})$} for the transitions between ordered and chaotic states. (\textbf{b}-\textbf{d}) Examples of the states identified in \textbf{a} with $-\frac{1}{2}$-defects (black) and $+\frac{1}{2}$--defects (white). Panel \textbf{d} highlights the antipolar ordering of $+\frac{1}{2}$-defect orientations (red bars); see also Fig.~\ref{fig:vortex} and Movie~S5 for a realization of this state in a $\sim10\times$ larger simulation domain.  
\label{fig:phase}
}
\end{figure}

\begin{figure*}[t!]
\includegraphics[width=0.8\textwidth]{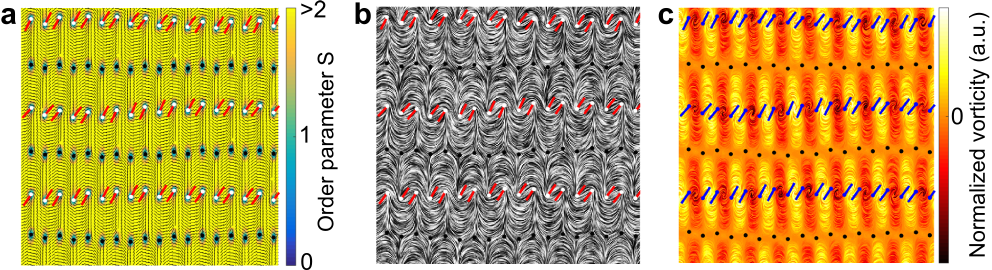}
\caption{Vortex lattice states with antipolar long-range ordering of nematic defects, see also Movie S5.
(\textbf{a}) Long-lived nematic order parameter field  with periodically aligned $-\frac{1}{2}$-defects (black) and $+\frac{1}{2}$-defects (red).
(\textbf{b}) Line integral convolution (LIC) plot of the corresponding director field as a proxy for microtubule-bundle patterns.
(\textbf{c}) LIC plot of the corresponding fluid velocity field, color-coded by normalized vorticity, demonstrates the formation of a vortex flow lattice. Dimensionless simulation parameters are $D=0.25,\;\gamma_2=1.875$.
\label{fig:vortex}
}
\end{figure*}

\begin{figure*}[ht]
\includegraphics[width=0.8\textwidth]{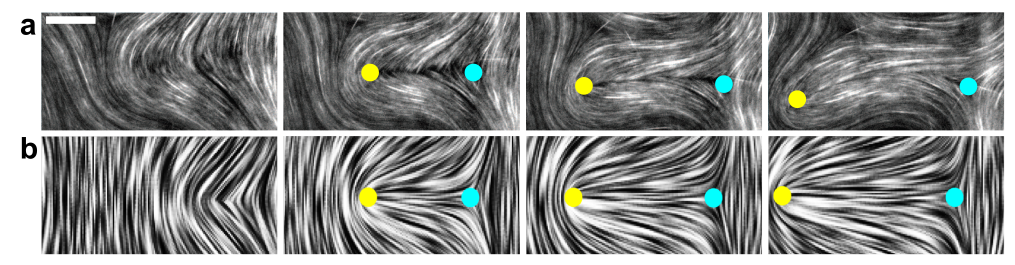}
\caption{Defect-pair creation and propagation in experiment and theory. (\textbf{a})  Experimentally observed dynamics of a defect pair, spontaneously produced by buckling and subsequent fracture of filaments; adapted with permission from Fig.~3\textbf{d} in Ref.~\cite{2012Sanchez_Nature}.  Scale bar $20\,\mu$m, time lapse 15$\,$s. (\textbf{b}) Line integral convolution (LIC) plot of the director fields showing the spontaneous creation and propagation of a defect-pair in a simulation of Eq.~\eqref{e:dimless} for~$D=1.5$, $\gamma_2=1$. As in the experiments, $+\f{1}{2}$-defects (yellow) generally move faster than $-\f{1}{2}$-defects (light blue), cf. Fig.~\ref{fig:exp_data}. 
\label{fig:pair}
}
\end{figure*}

\textbf{Phase diagram.}
To investigate the nonlinear dynamics of Eq.~\eqref{e:dimless}, we implemented a Fourier pseudospectral algorithm with modified Runge-Kutta time-stepping~\cite{Kassam} (Methods) and so evolved the real and imaginary parts of $\psi(t,z)$ in time for periodic boundary conditions in space.  A numerically obtained $(\gc_2,D)$-phase diagram for random initial conditions confirms the existence of a turbulent nematic phase if active self-advection is sufficiently strong (Fig.~\ref{fig:phase}\textbf{a},\textbf{b}; Movie~S1). Ordered configurations prevail at low activity (Fig.~\ref{fig:phase}\textbf{a},\textbf{c},\textbf{d};   Movies~S3, S4 and S6). Although the ground-states of the free energy \eqref{e:free-energy} are in general not homogeneous, the critical curve separating the two regimes is in fair agreement with the estimate  $D_c=2(-\gamma_2+\sqrt{\gamma_2^2+2})$ from linear stability of the homogenous state (white line in Fig.~\ref{fig:phase}\textbf{a}). For subcritical values of the advection parameter $D$, we observe either defect-free ground-states or long-lived lattice-like states exhibiting ordered defect configurations (Fig.~\ref{fig:phase}\textbf{c},\textbf{d}).  Regarding the subsequent comparison between theory and experiment, it is important to note that the lattices are also found in simulations with a large domain (Fig.~\ref{fig:vortex}; Movie~S5). These spatially periodic states generally exhibit antipolar long-range ordering of $+\f{1}{2}$-defects~(Fig.~\ref{fig:phase}\textbf{d}; Fig.~\ref{fig:vortex}\textbf{a}) accompanied by vortex flow lattices (Fig.~\ref{fig:vortex}\textbf{c}). Numerical free-energy calculations show that defect-free states (Fig.~\ref{fig:phase}\textbf{c}) typically have slightly lower energies than the lattice states (Fig.~\ref{fig:phase}\textbf{d}), leaving open the possibility of a very slow decay of the latter. However, regardless of whether such lattice states are extremely long-lived metastable or truly stable states, these simulation results confirm that antipolar ordering of $+\f{1}{2}$-defect pairs can persist over experimentally relevant time-scales.

\textbf{Theory vs. experiment.}
To test our theory systematically against existing experimental data~\cite{2012Sanchez_Nature,2015DeCamp}, we analyze defect-pair dynamics, global defect ordering and defect statistics in the turbulent nematic phase. 

\par
Spontaneous defect-pair creation and subsequent propagation, as reported in recent ALC experiments~\cite{2012Sanchez_Nature} and observed in our simulations,  are compared in Fig.~\ref{fig:pair}. In the experimental system~\cite{2012Sanchez_Nature}, a $(+\frac{1}{2},-\frac{1}{2})$-defect pair is created when fracture along incipient crack regions~\cite{PhysRevLett.114.048101} becomes energetically more favorable than buckling. After creation, the $+\frac{1}{2}$-defect moves away rapidly whereas the position of the $-\frac{1}{2}$-defect remains approximately fixed for up to several seconds (Fig.~\ref{fig:pair}\textbf{a}). \markblack{We note that an asymmetry in the speeds of topological defects has also been observed in passive liquid crystals \cite{2002Toth_PRL,2002Svensek_PRE,2005Blanc_PRL}.} Our simulations of the  minimal model defined in Eq.~\eqref{e:dimless} accurately reproduce the details of the experimentally observed dynamics (Fig.~\ref{fig:pair}\textbf{b}; Movies S1 and S2).  

\par
\begin{figure}[t]
\includegraphics[width=0.975\columnwidth]{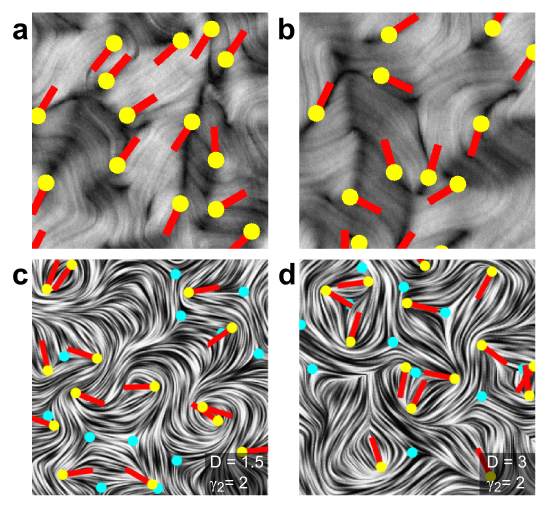}
\caption{Strong and weak antipolar ordering of $+\f{1}{2}$-defects as (\textbf{a},\textbf{b}) observed in experiments~\cite{2015DeCamp} and (\textbf{c},\textbf{d}) predicted by our theory based on 2D simulations with periodic boundary conditions.  Light-blue markers: $-\f{1}{2}$-defects. Yellow markers: $+\f{1}{2}$-defects. Red bars:  orientation of $+\f{1}{2}$-defects.   (\textbf{a})~$+\f{1}{2}$-defects in thin ALC films (thickness $h\sim 250\,$nm) show strong nematic alignment. (\textbf{b})~$+\f{1}{2}$-defects in thicker ALC films ($h\sim 1\,\mu$m) are more disordered. (\textbf{c},\textbf{d})  For weak effective hydrodynamic coupling $D$, simulations show antipolar ordering, which is inhibited for larger values of $D$. The average number of defects in the full simulation box is approximately (\textbf{c}) 240 and (\textbf{d}) 350. Figures \textbf{a} and \textbf{b} kindly provided by S. DeCamp and Z. Dogic.}
\label{fig:nematic_1}
\end{figure}
\begin{figure}[b]
\includegraphics[width=\columnwidth]{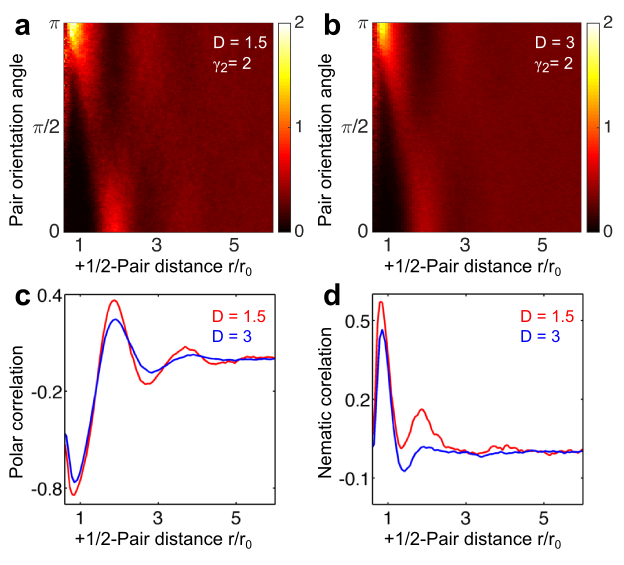}
\caption{ Increasing activity and film thickness decreases antipolar ordering in simulations.  (\textbf{a},\textbf{b}) Maxima of the numerically obtained local pair orientation PDFs $p(\theta_{ij}|r)$ signal antipolar local ordering of $+\f{1}{2}$-defects as they are separated by the typical defect-lattice spacing. The defect distance $r$ is specified in units of the mean nearest-neighbor distance $r_0$ between $+\f{1}{2}$-defects.  (\textbf{c},\textbf{d})~Polar $P(r)$ and nematic $N(r)$ correlation functions for $D=1.5$ (red) and $D=3$ (blue). Increasing the effective hydrodynamic coupling~$D$ leads to stronger mixing and hence decreases nematic order, which is corroborated by the nematic correlation length being $\sim40\%$ shorter for $D=3$ than for $D=1.5$ (panel \textbf{d}). This is also reflected by the diminished intensity of the maxima in panel \textbf{b} relative to panel~\textbf{a}. The simulation parameters correspond to those given in  Fig.~\ref{fig:nematic_1}\textbf{c},\textbf{d}.
\label{fig:nematic_2}}
\end{figure}

Another striking and unexplained experimental observation~\cite{2015DeCamp}  is the emergence of orientational order of $+\frac{1}{2}$-defects in thin ALC layers (Fig.~\ref{fig:nematic_1}). Using the setup illustrated  in Fig.~\ref{fig:closure}\textbf{b},  recent experiments~\cite{2015DeCamp} demonstrated nematic alignment of $+\frac{1}{2}$-defects in thin ALC layers of thickness $h \sim 250\,$nm  (Fig.~\ref{fig:nematic_1}\textbf{a}), whereas  thicker ALC layers with $h\sim 1\,\mu$m  showed no substantial orientational order on large scales (Fig.~\ref{fig:nematic_1}\textbf{b}). To investigate whether our theory can account for these phenomena, we tracked defect positions $\bs r_i$ (Methods) and defect orientations $\bs d_i=\nabla\cdot Q(\bs r_i)/|\nabla\cdot Q(\bs r_i)|$  \cite{2015Giomi_Arxiv} in simulations for different values of the advection parameter~$D=\zeta/\nu$, since Brinkman-type scaling arguments suggest that $D$ increases with the ALC layer thickness, $D\propto 1/\nu \propto h^p$ with $p\in[1,2]$.  
For weakly supercritical advection, $D\gtrsim D_c$, we find that Eq.~\eqref{e:dimless} predicts robust antipolar alignment of  $+\frac{1}{2}$-defects (Fig.~\ref{fig:nematic_1}\textbf{c}). In our simulations, this ordering decreases as the effective mixing strength $D$ increases  (Fig.~\ref{fig:nematic_1}\textbf{d}), consistent with the experimental results~\cite{2015DeCamp} for thicker  ALC layers (Fig.~\ref{fig:nematic_1}\textbf{b}). Similar ordering was observed in simulations that incorporated alignment of the nematic field near the horizontal boundaries of the simulation box (Supplementary Information; Fig.~S5). To quantify the degree of orientational order, we recorded the distances $d_{ij}$ between all $+\frac{1}{2}$-defect pairs $(i,j)$ as well as their relative orientation angles $\theta_{ij}=\cos^{-1}(\bs d_i\cdot \bs d_j)\in[0,\pi]$.  The resulting pair-orientation distributions $p(\theta|r)$, and polar and nematic correlation functions, $P(r) = \langle \bs d_i\cdot \bs d_j\rangle_r$ and $N(r)= 2\langle (\bs d_i\cdot \bs d_j)^2\rangle_r-1$, are shown in Fig.~\ref{fig:nematic_2}, with $\langle\cdot\rangle_r$ denoting an average over pairs of defects separated by a distance $r$. The local maxima in the orientation distribution at $\theta = 0$ and $\theta=\pi$ signal antipolar ordering (Fig.~\ref{fig:nematic_2}\textbf{a}), which is also reflected in the oscillatory behavior of the polar and nematic correlation functions  (Fig.~\ref{fig:nematic_2}\textbf{c},\textbf{d}). The diminished intensity of the local maxima for larger values of $D$  indicates that enhanced hydrodynamic mixing reduces orientational order (Fig.~\ref{fig:nematic_2}\textbf{c},\textbf{d}).

\begin{figure}[t!]
\includegraphics[width=0.95\columnwidth]{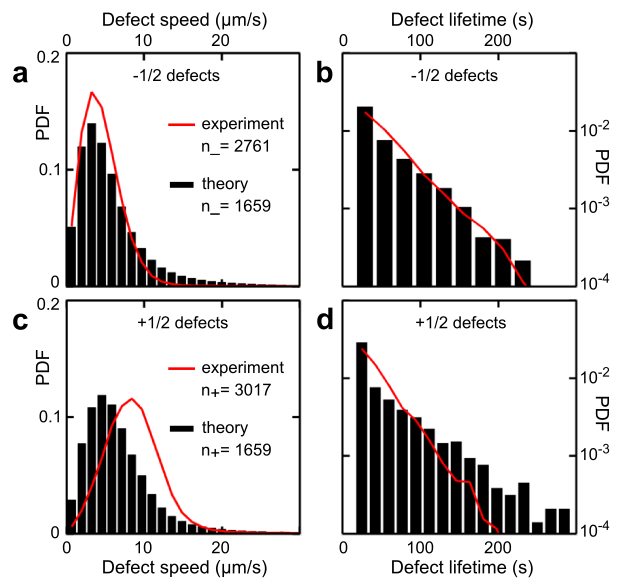}
\caption{Quantitative comparison of defect statistics between predictions of Eq.~\eqref{e:dimless} and   experimental data~\cite{2015DeCamp}, using the parameter estimation procedure described in the text. For $-\f{1}{2}$-defects, both (\textbf{a}) speed distribution and (\textbf{b}) lifetime distribution agree well. (\textbf{c}) For $+\f{1}{2}$-defects, experimentally measured speed values are slightly larger, as our model assumes a strongly overdamped limit. (\textbf{d})~Simulations with periodic boundary conditions (Movie~S8) predict a low-probability tail of large lifetimes which is not visible in the experiment, likely due to its restricted field of view or additional noise. Dimensionless simulation parameters $D=1.75$ and $\gamma_2=1$ translate into the following dimensional values:
$a = 0.08\,$s$^{-1}$, $b = 0.32\,$s$^{-1}$, $D = 1791\,\mu$m$^2$/s, $\gamma_2= 1024\,\mu$m$^2$/s, $\gamma_4 = 3.28\times 10^6\mu$m$^4$/s. The numbers $n_\pm$ reflect the number of $\pm\f{1}{2}$-defects tracked, and the simulation domain contained $\sim 130$ defects at any given time.
\label{fig:exp_data}
}
\end{figure}

\par
Lastly, we test the theoretically predicted defect statistics against a separate experimental data set kindly provided by DeCamp \textit{et al.} (private communication). Since our simulations are performed in dimensionless units, there is freedom to choose a characteristic lengthscale $l_0$ and  timescale $t_0$.  To relate theory and experiments, we determine $(l_0,t_0)$ such that the joint mean speed and mean lifetime of $\pm\f{1}{2}$-defects match the experimental values $\bar{v}=6.6\,\mu$m/s and $\bar{\tau}=52.8\,$s. After fixing these global scales, we can compare details of the speed and lifetime distributions  (Fig.~\ref{fig:exp_data}). To this end,  we first locate the \lq best-fit\rq{} simulation parameters in the $(\gamma_2,D)$-parameter space explored in the phase diagram (Fig.~\ref{fig:phase}\textbf{a}). This procedure identifies  $\gamma_2=1,D=1.75$ as the best-match parameters, although nearby parameter values and simulations with $\kappa=1$ produce fits of similar quality, corroborating the robustness of the model (Fig.~S4). For $-\f{1}{2}$-defects, we find adequate agreement between experiment and theory for speed and lifetime probability density functions (PDFs), as evident from Fig.~\ref{fig:exp_data}\textbf{a,b}. For $+\f{1}{2}$-defects, simulation results also agree well with the experimental measurements  (Fig.~\ref{fig:exp_data}\textbf{c},\textbf{d}), but one notices two systematic differences. First, while the peak heights of the PDFs agree within a few percent,   experimentally measured speed values for $+\f{1}{2}$-defects are on average slightly larger than theoretically predicted values (Fig.~\ref{fig:exp_data}\textbf{c}). Second, simulation data predict a miniscule tail-fraction of long-living $+\f{1}{2}$-defects  not detected in the experiment (Fig.~\ref{fig:exp_data}\textbf{d}).  In addition, based on the experimental density estimate of 30 defects/mm$^2$~\cite{2015DeCamp}, we find that the  defect density at any given time in the \lq{}best-fit\rq{} simulation is $\sim2.3\times$ lower than in the experiments. As discussed below, such deviations can be explained plausibly by specific model assumptions. Taken together, the above results  confirm that the minimal model defined by Eq.~\eqref{e:dimless} provides a satisfactory qualitative and quantitative description of the main experimental results~\cite{2012Sanchez_Nature,2015DeCamp}.

\section*{DISCUSSION}

\textbf{Pattern-formation mechanism.}
Equations~\eqref{e:free-energy} and~\eqref{e:free_energy-complex} epitomize the idea of \lq universality\rq{} in spatio-temporal pattern formation, as known from Swift-Hohenberg-type scalar field theories~\cite{1977SwiftHohenberg,2006Aranson}. The free-energy expressions contain the leading-order terms of generic series expansions in both order-parameter space and Fourier space, consistent with spatial and nematic symmetries. When considering passive systems  with a preference for  homogenization ($\gc_2<0$), it usually suffices to keep only the quadratic gradient terms.  By contrast, for pattern forming systems, the coefficient in front of the lowest-order gradient contribution can change sign~\cite{1977SwiftHohenberg,2015Stoop_NatMat}, and one must include higher-order derivatives to ensure stability. We hypothesize that the sign change of $\gamma_2$ is directly related to the motor-induced buckling of microtubule bundles (Fig.~\ref{fig:closure}\textbf{a}), an effect that is not captured by the standard LdG free-energy for passive liquid crystals. In a few select cases, expressions of the form~\eqref{e:free-energy} and~\eqref{e:free_energy-complex} can be systematically derived~\cite{1977SwiftHohenberg,2014Grossmann_PRL,2015Stoop_NatMat}. Generally, one can regard the free-energy expansion~\eqref{e:free_energy-complex}  as an effective field theory whose phenomenological parameters can be determined from experiments. This  approach has proved successful for dense bacterial suspensions~\cite{2012Wensink,2013Dunkel_PRL}  and now also for ALCs, indeed suggesting some universality in the formation and dynamics of topological defects in active systems.

\textbf{Nematic defect order.}
Although $(\gc_2,D)$ are varied as independent effective parameters in the simulations, they are likely coupled through underlying physical and chemical parameters. For example, it is plausible that a change in ATP-concentration or film-thickness would affect both~$\gamma_2$ and $D$.  The parameter $D$ can also be interpreted as an effective Reynolds number. In our numerical exploration of the $(\gamma_2,D)$-parameter space, we observe  for subcritical advection $D$ either long-lived lattice-like states exhibiting nematically aligned $+\f{1}{2}$-defects or defect-free ground-states (Figs.~\ref{fig:phase},~\ref{fig:vortex}; Movies~S3, S4, S5, S6 and S7). Ordered defect configurations correspond to local minima or saddles in the free-energy landscape and have only slightly higher energy than defect-free states (Fig.~\ref{fig:phase}\textbf{c}). When the activity $\zeta$ is sufficiently large that advection is marginally supercritical, $D\gtrsim D_c$, chaotic system trajectories spend a considerable time in the vicinity of these metastable lattice states, which provides a physical basis for the orientational order of defects in thin ALC films~\cite{2015DeCamp}. For $D\gg D_c$, the ALC system can access a wider range of high-energy states, leading to increased disorder in the defect dynamics. Although the strongly turbulent regime $D\gg D_c$ requires high time-resolution and is thus difficult to realize in long-time simulations, the inhibition of nematic defect order at larger values of $D$ is evident from the reduced peak heights in Fig.~\ref{fig:nematic_2}\textbf{c} and \textbf{d}.

\textbf{Defect statistics.}
The systematic speed deficit  in Fig.~\ref{fig:exp_data}\textbf{c} likely reflects the overdamped closure condition~\eqref{e:closure}, which suppresses the propagation of hydrodynamic excitations. Since flow in a newly created defect pair generally points from the $-\f{1}{2}$ to the $+\f{1}{2}$-defect, the minimal model~\eqref{e:dimless} can be expected to underestimate the speeds of $+\f{1}{2}$-defects. This effect could be explored in future experiments through a controlled variation of the thickness and viscosity of the oil film (Fig.~\ref{fig:closure}\textbf{b}).  The low-probability tail  of long-living $+\f{1}{2}$-defects in the simulation data (Fig.~\ref{fig:exp_data}\textbf{d}) may be due to the fact that  they can be tracked indefinitely in the simulations but are likely to  leave the finite field of view in the experiments. In the future, the minimal theory presented here should be extended systematically by adding physically permissible extra terms~\cite{2003StarkLubensky} to the free-energy, explicitly simulating the full 
hydrodynamics in Eq.~\eqref{e:Stokes}, or incorporating additional terms into Eq.~(\ref{e:Q-dynamics}) that account for the interaction between the nematic field and the induced flow~\cite{2003StarkLubensky,2013Giomi_PRL}. 

\markblack{\textbf{Future extensions.}
The minimal model formulated in Eqs.~(\ref{e:Q-dynamics}) and (\ref{e:free-energy}) can be systematically extended to improve further the quantitative agreement between experiment and theory. For instance, one may append to the right-hand side of (\ref{e:Q-dynamics}) an additional fourth-order linear term of the form $(\nabla\nabla)^+\left[\nabla\cdot(\nabla\cdot Q)\right]$ \cite{2013Bertin_NJP}, $\mathcal{O}^+$ denoting the symmetric traceless part of the operator $\mathcal{O}$. Such a term only affects the high-wavenumber damping at order $k^4$ and thus is not expected to alter significantly  the results obtained here. We also note that extra terms coupling the nematic field to the induced flow may be added to Eq.~\eqref{e:Q-dynamics}, an example being $SE$, where $E=(1/2)\left[\nabla \bs v+(\nabla \bs v)^\top\right]$ is the symmetrized strain rate tensor \cite{2016Putzig_SoftMatter,2013Giomi_PRL}. Our above analysis neglected such secondary hydrodynamic effects in the interest of constructing a minimal  mathematical theory capable of capturing key experimental observations. Moreover, this simplification may be justified on the physical basis that steric interactions and motor-induced buckling are expected to dominate over flow-alignment effects at high microtubule densities\footnote{In active nematics of low to intermediate density, concentration fluctuations can trigger additional instabilities~\cite{PhysRevLett.113.038302}.}. The effect of microtubule bending may be enhanced by appending a hydrodynamic interaction term proportional to $E^+$ \cite{2016Putzig_SoftMatter}, since the closure condition $\bs v = -D\nabla\cdot Q$ implies that $E^+ = -({D}/2)\Delta Q$, which augments the bending term $-\gamma_2\Delta Q$ in Eq.~\eqref{e:free-energy}. \markblue{However, the scale separation between the experimentally observed filament  buckling wavelength and the flow structures in the isotropic phase at low microtubule concentrations (see Fig.~1d in Ref.~\cite{2012Sanchez_Nature}) suggests that such hydrodynamic effects play a secondary role}. A natural next step would be to derive systematically the bending term from a microscopic model of motors and filaments as introduced in Ref.~\cite{PhysRevLett.114.048101}. Finally, it will be worthwhile to attempt constructing a fully 3D theory for the ALCs and fluid and subsequently project on the quasi-2D interface, although additional assumptions are required then to obtain a closed 2D system of equations.}

\section*{CONCLUSIONS}

Recent experimental and theoretical studies showed that fourth-order PDE models for scalar and vector fields provide an accurate quantitative description of surface-pattern formation in soft elastic materials~\cite{2015Stoop_NatMat} and orientational order in dense bacterial fluids \cite{2012Wensink,2013Dunkel_PRL}. Here, we have generalized these ideas to  matrix-valued fields describing soft active nematics. The above analysis demonstrates that a generic fourth-order $Q$-tensor model can shed light on experimental observations in 2D ALCs~\cite{2015DeCamp}, including the emergence of orientational order of topological defects. Physically, the higher-order generalization~\eqref{e:free-energy} becomes necessary because the commonly adopted LdG free-energy, which was designed to describe passive liquid crystals, does not account for the experimentally observed spontaneous buckling of motor-driven ALCs~\cite{2012Sanchez_Nature}.  More generally, the fact that three vastly different soft matter systems can be treated quantitatively in terms of structurally similar higher-order PDEs~\cite{2012Wensink,2013Dunkel_PRL,2015Stoop_NatMat} promises a unified mathematical framework for the description of pattern formation processes in a broad class of complex materials. In addition, the free-energy analogy~\cite{PhysRevA.38.2132}  between dense ALCs and generalized Gross-Pitaevskii models  suggests that the self-organization principles~\cite{Goldenfeld:2011aa}  of mesoscopic active matter and microscopic quantum systems~\cite{PhysRevLett.88.090401,Lin:2011aa,PhysRevLett.109.095302,Parker:2013aa}  could be more similar than previously thought.


\textbf{Acknowledgments.}
This work was supported by the NSF Mathematical Sciences Postdoctoral Research Fellowship DMS-1400934 (A.O.), an MIT Solomon Buchsbaum Fund Award  (J.D.) and an Alfred P. Sloan Research Fellowship (J.D.). The authors would like to thank Zvonimir Dogic and  Stephen DeCamp for sharing their experimental data and explaining the details of their experiments, and Hugues Chat\'e, Michael Hagan, Jean-Francois Joanny, Ken Kamrin, Mehran Kardar, Cristina Marchetti, Jonasz Slomka, Norbert Stoop, Francis Woodhouse and  Martin Zwierlein for insightful discussions and helpful comments.
\vspace{-0.4cm}

\subsection*{METHODS}
\vspace{-0.3cm}
\textbf{Numerical solver \& defect tracking.}
To simulate Eq.~\eqref{e:dimless}, we implemented a numerical algorithm that evolves the real and imaginary parts $\gl(t,\bs r)$ and $\mu(t,\bs r)$ of the complex order parameter $\psi$ in time for periodic boundary conditions in space. The algorithm solves Eq.~\eqref{e:dimless} pseudospectrally in space using $N_\ell=256$ or $N_\ell=512$ lattice points in each direction and a simulation box of size $L = 6\pi$ (Fig. 2 and 4) or $L = 18\pi$ (Fig. 3, 5-7). Spectral analysis shows that $N_\ell =256$ is generally sufficient to resolve the fine-structure of the numerical solutions~(Supplementary Fig. S6). The algorithm steps forward in time using a modified exponential time-differencing fourth-order Runge-Kutta method~\cite{Kassam} with time step $\Delta t \leq 2^{-10}$. Simulations were initialized with either a single defect pair or random field configurations $\{\lambda(0,\bs r),\mu(0,\bs r)\}$. Defects are located at the intersections of the zero-contours of $\gl$ and $\mu$, their positions tracked by implementing James Munkres' variant of the Hungarian assignment algorithm~\cite{Munkres} (Supplementary Information).


\end{document}